\documentclass[11pt]{article}
\usepackage{comment}
\usepackage[margin=2.5cm]{geometry}
\usepackage[toc,page]{appendix}
\usepackage{amsmath,amsfonts,breqn,comment,cancel,mathtools,amssymb,extarrows,mathtools,graphicx,subfigure,setspace,bbold,physics}
\usepackage{cite}
\usepackage{slashed}
\usepackage[dvipsnames]{xcolor}
\usepackage{hyperref}
\hypersetup{colorlinks=true, linkcolor=blue, citecolor=magenta, linktoc=page}
\usepackage{url}
\usepackage{tikz}
\usepackage{color}
\usepackage{arydshln,leftidx,mathtools}
\usetikzlibrary{decorations.pathreplacing}
\numberwithin{equation}{section}
\newcommand{\be}{\begin{equation}}
	\newcommand{\bea}{\begin{eqnarray}}
		\newcommand{\eea}{\end{eqnarray}}
	\newcommand{\ba}{\begin{align}}
		\newcommand{\ea}{\end{align}}
	\newcommand{\ee}{\end{equation}}

\usepackage{chngcntr}
\begin{document}
	\onehalfspacing
\begin{titlepage}
	\thispagestyle{empty}

%
	
	\vspace{.4cm}
	\begin{center}
		\noindent{\Large \textbf{Capacity of Entanglement in Lifshitz Theories}}\\
		
		\vspace*{15mm}
	\end{center}
	\begin{center}
{Sare Khoshdooni,  Komeil Babaei Velni and M. Reza Mohammadi Mozaffar

}
\end{center}
\vspace*{0.1cm}

\begin{center}
{\it Department of Physics, University of Guilan, P.O. Box 41335-1914, Rasht, Iran		}
		
		\vspace*{0.5cm}
		{E-mails: {\tt\{sare.khoshdooni@webmail., babaeivelni@, mmohammadi@\}guilan.ac.ir}}%
\end{center}
	\begin{abstract}
We study the capacity of entanglement in certain integrable scale invariant theories that exhibit Lifshitz scaling
symmetry with a generic dynamical exponent $z$ at the critical point. This measure characterizes the width of the eigenvalue spectrum of the reduced density matrix and is a quantum informational counterpart of heat capacity. We explore various aspects of capacity of entanglement, such as the corresponding universal terms for the ground state, its $z$-dependence, and its temporal evolution, after a global quantum quench in two dimensions. We carefully examine the existence of a convenient entropic $c$-function based on this quantity both in bosonic and fermionic theories. While in the relativistic case the corresponding $c$-function displays a monotonic behavior under the renormalization group flow, this is not the case for the nonrelativistic theories. We also investigate the dynamics of capacity of entanglement after a mass quench and show that it follows the quasiparticle interpretation for the spreading of entanglement. Finally, we discuss how these results are consistent with the behavior of other entanglement measures including the entanglement entropy.

\end{abstract}

\end{titlepage}
\newpage

\tableofcontents
\noindent
\hrulefill


\section{Introduction}\label{intro}

Quantum entanglement has been increasingly studied over the last decades in a variety of areas of physics, including condensed matter physics, quantum field theory and quantum gravity (e.g., see Refs. \cite{Horodecki:2009zz,Calabrese:2009qy,Nishioka:2018khk,Casini:2022rlv} as reviews). In this context, different entanglement and information measures have proved very useful for developing our understanding of the quantum correlations, e.g., entanglement entropy and logarithmic negativity. Indeed, for a system described by a pure state, the entanglement entropy is the unique measure that quantifies the amount of quantum entanglement between two complement subsystems. In order to define this
measure, let us consider a bipartite system in a pure state $|\psi\rangle$ with Hilbert space equal to the direct product of two factors, i.e., $\mathcal{H}=\mathcal{H}_A\otimes \mathcal{H}_{\bar{A}}$. By summing over the degrees of freedom in $\mathcal{H}_{\bar{A}}$ we can find the reduced density matrix for $A$, i.e., $\rho_A={\rm Tr}_{\bar{A}}\;|\psi\rangle\langle\psi|$. Since the reduced density matrix is both Hermitian and positive semidefinite, it can be expressed as $\rho_A=e^{-H_A}$, where $H_A$ is the modular Hamiltonian. Now the entanglement entropy would then be the von Neumann entropy of $\rho_A$,
\begin{eqnarray}\label{EE}
S_{E}=-{\rm Tr}_A\left(\rho_A \log \rho_A\right)=\langle H_A\rangle.
\end{eqnarray}
Thus, we can consider the entanglement entropy as the expectation value of the modular Hamiltonian. Further, to gain more information about the spectrum of $\rho_A$ we can consider the second cumulant (variance) of the modular Hamiltonian, which is called the capacity of entanglement\cite{Yao:2010woi}
\begin{eqnarray}\label{capa}
C_E=\langle H_A^2 \rangle-\langle H_A \rangle^2.
\end{eqnarray}
Hence, this measure characterizes the width of the eigenvalue distribution of $\rho_A$ such that for a maximally entangled state, it vanishes. Unfortunately, the above measures are notoriously difficult to compute, mainly because one does not have a good way to find the logarithm of an operator. Indeed, one can obtain the entanglement entropy and capacity of entanglement indirectly using the Renyi entropy, defined as
\begin{eqnarray}\label{renyi}
S_{n}=\frac{1}{1-n}\log {\rm Tr}\rho_A^n,
\end{eqnarray}
where $n$ is a positive integer. Using the above expression, it is straightforward to show that 
\begin{eqnarray}\label{SECE}
S_E=\lim_{n\rightarrow 1} S_n, \hspace{2cm} C_E\equiv \lim_{n\rightarrow 1}C_n=\lim_{n\rightarrow 1} n^2\frac{\partial^2}{\partial n^2}\left((1-n)S_n\right),
\end{eqnarray}
where $C_n$ is the $n$th capacity of entanglement. Recently, there have been a wide variety of research efforts investigating the properties of the capacity of entanglement in different setups both in the field theories and holography \cite{DeBoer:2018kvc,Nakaguchi:2016zqi,Nakagawa:2017wis,Kawabata:2021hac,Okuyama:2021ylc,Kawabata:2021vyo,Nandy:2021hmk,Bhattacharjee:2021jff,Wei:2022bed,Shrimali:2022bvt,Arias:2023kni,Andrzejewski:2023dja,Ren:2024qmx,Banks:2024cqo,Shrimali:2024nbc,MohammadiMozaffar:2024uyp,MohammadiMozaffar:2024uiy,Andrzejewski:2025}. In particular, considering the above measures, an interesting observation in \cite{DeBoer:2018kvc} was that the entanglement entropy and capacity of entanglement coincide with each other in certain setups, e.g., $(1+1)$-dimensional conformal field theories at equilibrium. Further, as examined in this reference, this intriguing behavior also arises in field theories with a holographic bulk gravity dual, where $C_E$ is comparable to $S_E$ and obeys an area law divergence. Another important observation is that a specific $c$-function, which is defined based on the capacity of entanglement, displays a monotonic behavior under the renormalization group (RG) flow\cite{Arias:2023kni}.

An interesting question to ask is whether the above results also hold in more general setups such as nonrelativistic quantum field theories. 
Indeed, our goal in this paper is to investigate the behavior of capacity of entanglement in integrable theories in which the scale invariance is anisotropic between time and space, the so-called Lifshitz quantum field theories.  As is well known, a physical system typically exhibits a Lifshitz scaling symmetry at quantum critical points. A Lifshitz invariant theory admits the nonrelativistic scaling symmetry
\begin{eqnarray}\label{lifshitzscaling1}
t\rightarrow \lambda^z \,t,\hspace*{2cm}x_i\rightarrow \lambda \,x_i,
\end{eqnarray}
where $z$ denotes the dynamical exponent. In this paper, we aim to provide a detailed study of the influence of anisotropic scale invariance on the behavior of capacity of entanglement. Toward this goal, we consider two families of bosonic and fermionic theories that have anisotropic scale invariant UV fixed points. Let us recall that quantum entanglement has already been extensively studied in the context of nonrelativistic models to  investigate the universal behaviors of entanglement measures both in static and time-dependent setups, e.g., \cite{Solodukhin:2009sk,Nesterov:2010yi,Keranen:2011xs,Mintchev:2022xqh,Mintchev:2022yuo}. Moreover, some investigations attempting to better understand the behavior of different information measures in Lifshitz field theories have also appeared \cite{Fradkin:2006mb,Zhou:2016ykv,MohammadiMozaffar:2017nri,He:2017wla,MohammadiMozaffar:2017chk,Angel-Ramelli:2020wfo,Mollabashi:2020ifv,Boudreault:2021pgj,Berthiere:2023bwn,Basak:2023otu,Vasli:2024mrf,Vasli:2023syq}. Despite extensive studies of different aspects of entanglement measures in theories with Lifshitz scaling symmetry, the behavior of capacity of entanglement has not been investigated. Hence, in this paper, we explore the behavior of the entanglement spectrum and capacity of entanglement in such theories. Moreover, we will discuss how our results are comparable with the behavior of other measures, including Renyi entropies. We also investigate the existence of consistent entanglement $c$-functions based on the capacity of entanglement.

The rest of the paper is organized as follows: In Sec. \ref{sec:setup}, we give the general framework in which we are working, establishing our notation and the general form of the bosonic and fermionic models whose actions respect Lifshitz scaling symmetry at the UV fixed point. In Sec. \ref{sec:static}, we evaluate the capacity of entanglement for ground state of a family of Lifshitz field theories numerically. To get a better understanding of the results, we will
also compare the behavior of this quantity to other correlation measures, including entanglement and Renyi entropies. 
In Sec. \ref{sec:cfunc}, we explore whether an entanglement candidate $c$-function based on capacity of entanglement can exhibit
monotonicity along the renormalization group flow generated by the mass. In Sec. \ref{sec:quench}, we investigate the time evolution of capacity of entanglement after a global quantum quench, where we present both numerical and semianalytic results. We conclude with a discussion of our results, as well as possible future directions, in Sec. \ref{sec:diss}.

\section{Setup} \label{sec:setup}

In this section, we briefly review some preliminaries to introduce the integrable theories in which the scale invariance is anisotropic between time and space. We also review the correlator method which is an efficient algorithm to compute the eigenvalues of the reduced density matrix in general Gaussian states both in static and time-dependent cases. 
For simplicity in the rest of the discussion, we focus our analysis on the special case of two-dimensional field theories. It is worth noting that the main interesting qualitative features of the capacity of entanglement are independent of the dimensionality of the theory.

\subsection*{Bosonic Lifshitz theory}

The first model we consider is a free-scalar theory in $1+1$ dimensions whose action is given by
\begin{eqnarray}\label{actionscalar}
I_b=\frac{1}{2}\int dt\,dx\;\left[(\partial_t\phi)^2-(\partial_x^z\phi)^2-m^{2z}\phi^2\right],
\end{eqnarray}
which is invariant under the scaling introduced in Eq. \eqref{lifshitzscaling1} in the massless limit. Note that the mass dimension of the scalar field is $(1-z)/2$, which explicitly depends on the dynamical exponent. Several aspects of entanglement measures for static situations in this theory have been studied in \cite{MohammadiMozaffar:2017nri,He:2017wla,MohammadiMozaffar:2017chk,Mollabashi:2020ifv}. In the next sections, much of our discussion will focus on the different aspects of capacity of entanglement in both static and time-dependent setups to examine how a nontrivial dynamical exponent can affect different scaling behaviors. Moreover, to obtain finite results for the entanglement measures, we regulate the above theory by placing it on a one-dimensional lattice. Indeed, considering a lattice with $N$ sites, the corresponding discrete Hamiltonian density is given by
\begin{eqnarray}\label{Hamildisscalar}
\mathcal{H}_b=\frac{1}{2}\sum_{n=1}^{N}\left(p_n^2+\left(\sum_{j=0}^z(-1)^{z+j} \frac{z!}{j!(z-j)!}q_{n-1+j}\right)+m^{2z}q_n^2\right),
\end{eqnarray}
where without loss of generality we have fixed the lattice spacing to unity. Further, in the momentum basis, the above expression can be rewritten as follows \cite{MohammadiMozaffar:2017nri}:
\begin{eqnarray}\label{Hamildissscalar1}
\mathcal{H}_b=\int_{-\pi}^{\pi}dk\,\omega_{k}\left(a^\dagger_{-k}a_{-k}+a^\dagger_{k}a_{k}\right),
\end{eqnarray}
where 
\begin{eqnarray}\label{dipersionscalar}
\omega_{k}=\sqrt{m^{2z}+\left(2\sin \frac{k}{2}\right)^{2z}}
\end{eqnarray}
is the corresponding dispersion relation. In order to find the spectrum of the reduced density matrix and the corresponding entanglement measures in a general Gaussian state, we use the correlator method \cite{Peschel:2002yqj,Casini:2009sr,Eisler:2009vye}. Indeed, considering a periodic lattice in the static vacuum state we do so by first constructing the following correlation functions:
\begin{eqnarray}\label{scalarcor}
X\equiv\langle q_r\,q_s\rangle=\frac{1}{2\pi}\int_{-\pi}^{\pi}\frac{e^{i(r-s)k}}{\omega_k}\,dk,\hspace{1.5cm}P\equiv\langle p_r\,p_s\rangle=\frac{1}{2\pi}\int_{-\pi}^{\pi}\omega_k \,e^{i(r-s)k}\,dk,
\end{eqnarray}
where $r, s\in [1, \ell]$, and $\ell$ denotes the number of lattice points enclosed by the entangling subregion. Now, defining the matrix $(X.P)^{1/2}$ whose eigenvalues are $\{\xi_j\}$, the corresponding expression for the Renyi entropy becomes
\begin{eqnarray}\label{Renyi}
S_n=\frac{1}{n-1}\sum_{j=1}^{\ell}\log\left(\left(\xi_j+\frac{1}{2}\right)^n-\left(\xi_j-\frac{1}{2}\right)^n\right).
\end{eqnarray}
Next, using Eq. \eqref{SECE} we can also find the entanglement entropy and capacity of entanglement, as follows:
\begin{eqnarray}
S_E&=&\sum_{j=1}^{\ell}\left(\left(\xi_j+\frac{1}{2}\right)\log\left(\xi_j+\frac{1}{2}\right)-\left(\xi_j-\frac{1}{2}\right)\log\left(\xi_j-\frac{1}{2}\right)\right),\label{SE}\\
C_E&=&\sum_{j=1}^{\ell}\left(\xi_j^2-\frac{1}{4}\right)\left(\log\frac{\xi_j-\frac{1}{2}}{\xi_j+\frac{1}{2}}\right)^2\label{CE}.
\end{eqnarray}
Before we proceed, let us recall that in the massless limit, the correlators defined in Eq. \eqref{scalarcor} diverge, which is due to the existence of a zero mode with vanishingly small momentum. Hence, for periodic boundary conditions we work at finite but small $m$, such that $m\ell\ll 1$. As is well known, another way to get rid of this zero mode is to break the translational symmetry, e.g., replacing the periodic boundary condition with the
Dirichlet boundary condition. In this case the corresponding correlators in the thermodynamic limit become \cite{MohammadiMozaffar:2017nri}
\begin{eqnarray}\label{scalarcorDir}
X=\frac{1}{2\pi}\int_{-\pi}^{\pi}\frac{\sin(rk)\sin(sk)}{\tilde{\omega}_k}\,dk,\hspace{1.5cm}P=\frac{1}{2\pi}\int_{-\pi}^{\pi}\tilde{\omega}_k \sin(rk)\sin(sk)\,dk,
\end{eqnarray}
where $\tilde{\omega}_k=\omega_{k/2}$. Also, the measures can be computed in a similar manner as before, which is straightforward.

In order to study the dynamics of entanglement measures one can generalize this approach to time-dependent cases, e.g., quantum quenches from a massive theory to a massless (scale invariant) theory. To do so, we consider a quantum system in the ground state $|\psi\rangle_0$ of the Hamiltonian $H_0$. Then, at $t=0$ one changes some parameter(s) of the system and modifies the Hamiltonian, i.e., $H_0\rightarrow H$. In general, $[H_0, H]\neq 0$ and thus $|\psi\rangle_0$ becomes an excited state with respect to the new Hamiltonian. Indeed, the evolution of the initial state toward the equilibrium is given by $|\psi(t)\rangle=e^{-iHt}|\psi\rangle_0$, which is highly nontrivial. In this case, the reduced density matrix is $\rho_A(t)={\rm Tr}_{\bar{A}}\;|\psi(t)\rangle\langle\psi(t)|$, hence the corresponding entanglement measures become time-dependent. In what follows, we study the temporal evolution of the entanglement measures after the global quench in Lifshitz theories, where the mass parameter is suddenly changed from $m_0$ to $m$.
It is straightforward to show that in this case, the corresponding time-dependent correlators become \cite{Coser:2014gsa,Cotler:2016acd}
\begin{eqnarray}
&&\mathbb{Q}_{r,s}(t)\equiv \langle \psi_{0}|q_{r}(t)q_{s}(t)|\psi_{0}\rangle=\frac{1}{2\pi}\int_{-\pi}^{\pi}Q_{k}(t)e^{i(r-s)k},\nonumber\\
&&\mathbb{P}_{r,s}(t)\equiv \langle \psi_{0}|p_{r}(t)p_{s}(t)|\psi_{0}\rangle=\frac{1}{2\pi}\int_{-\pi}^{\pi}P_{k}(t)e^{i(r-s)k},\\
&&\mathbb{M}_{r,s}(t)\equiv \langle \psi_{0}|\frac{1}{2}\{q_{r}(t),p_{s}(t)\}|\psi_{0}\rangle=\frac{1}{2\pi}\int_{-\pi}^{\pi}M_{k}(t)e^{i(r-s)k},\nonumber
\end{eqnarray}
where 
\begin{eqnarray}
&&Q_{k}(t)\equiv\frac{1}{\omega_{k}}(\frac{\omega_{k}}{\omega_{0,k}}\cos[2](\omega_{k}t)+\frac{\omega_{0,k}}{\omega_{k}}\sin[2](\omega_{k}t)),\nonumber\\
&&P_{k}(t)\equiv \omega_{k}(\frac{\omega_{k}}{\omega_{0,k}}\sin[2](\omega_{k}t)+\frac{\omega_{0,k}}{\omega_{k}}\cos[2](\omega_{k}t)),\\
&&M_{k}(t)\equiv(\frac{\omega_{k}}{\omega_{0,k}}+\frac{\omega_{0,k}}{\omega_{k}})\sin(\omega_{k}t)\cos(\omega_{k}t),\nonumber
\end{eqnarray}
and $\omega_{0,k}$ is given by Eq. \eqref{dipersionscalar} with $m=m_0$. Moreover, the corresponding measures are obtained by the replacement $\xi_j\rightarrow \xi_j(t)$ in Eqs. \eqref{SE} and \eqref{CE}. Here, $\{\xi_j(t); 1\leq j\leq \ell\}$ with $\xi_j(t)>0$ can be determined by evaluating the spectrum of $iJ_A.\Gamma_A(t)$, where 
\begin{equation}
	\Gamma_A(t)={\rm Re}\begin{pmatrix}
		\mathbb{Q}_{A}(t)&\mathbb{M}_{A}(t)\\
		\mathbb{M}_{A}^{T}(t)&\mathbb{P}_{A}(t)\\
	\end{pmatrix}, \qquad 
	J_A=\begin{pmatrix}
		\bold{0}_{\ell\times\ell}&\bold{1}_{\ell\times\ell}\\
		\bold{-1}_{\ell\times\ell}&\bold{0}_{\ell\times\ell}\\
	\end{pmatrix}.
\end{equation}

\subsection*{Fermionic Lifshitz theory}

The second model we consider is a two-dimensional free-fermion theory whose action is given by
\begin{eqnarray}\label{actiondirac}
I_f=\frac{1}{2}\int dt\,dx\;\Psi\left[i\gamma^t\partial_t+i\gamma^xT^{z-1}\partial_x-m^{z}\right]\Psi,
\end{eqnarray}
where $T=\sqrt{-\partial_x\partial^x}$. In this case, the mass dimension of $\Psi$ is $\frac{1}{2}$,, which is independent of the dynamical exponent. The corresponding entanglement measures have been studied in \cite{Vasli:2024mrf} (see also \cite{Hartmann:2021vrt} for other developments on the general properties of the entanglement measures in slightly different fermionic Lifshitz models). Again, the above model is invariant under the scaling introduced in Eq. \eqref{lifshitzscaling1} in the massless limit. Further, the corresponding discretized Hamiltonian is given by
\begin{eqnarray}\label{Hamildisdirac}
\mathcal{H}_f=\sum_{n=1}^{N}\Psi_n^{\dagger}\left(-\gamma^2 f(k)+m^z\gamma^0\right)\Psi_n,
\end{eqnarray}
where the expression of $f(k)$ for odd and even values of the dynamical exponent is
\begin{eqnarray}\label{fdirac}
f_o(k)=-\left(\sin k\right)^z,\hspace{2cm}f_e(k)=|\sin k|\,\left(\sin k\right)^{z-1}.
\end{eqnarray}
Moreover, in the momentum basis the Hamiltonian becomes
\begin{eqnarray}\label{Hamildissdirac1}
\mathcal{H}_f=\int_{-\pi}^{\pi}dk\,\omega_{k}\left(a^\dagger_{-k}a_{-k}+a^\dagger_{k}a_{k}\right),
\end{eqnarray}
where
\begin{eqnarray}\label{dispersionfermion}
\omega_{k}=\sqrt{m^{2z}+f^2(k)}.
\end{eqnarray}
Again we employ the correlator method to find the the spectrum of the reduced density matrix and the corresponding entanglement measures. Indeed, in this case the fermionic two-point function for the  vacuum state is given by \cite{Vasli:2024mrf}
\begin{eqnarray}\label{diraccor}
C_{rs}\equiv\langle \Psi_r^\dagger \Psi_s \rangle=\frac{\delta_{rs}}{2}\textbf{1}_{2\times 2}-\frac{1}{4\pi}\int_{-\pi}^{\pi}\frac{dk}{\omega_k}
\left(\begin{matrix}
f(k) & m^z\\
m^z & -f(k)
\end{matrix}\right) e^{i(r-s)k}.
\end{eqnarray}
Now, denoting the eigenvalues of the above matrix by $\{\xi_j\}$, for the Renyi entropy we have
\begin{eqnarray}\label{Renyidirac}
S_n=\frac{1}{1-n}\sum_{j=1}^{\ell}\log\left(\xi_j^n+\left(1-\xi_j\right)^n\right).
\end{eqnarray}
Further, using Eq. \eqref{SECE}, the corresponding expressions for the entanglement entropy and capacity of entanglement are
\begin{eqnarray}
S_E&=&-\sum_{j=1}^{\ell}\left(\xi_j\log\xi_j+\left(1-\xi_j\right)\log\left(1-\xi_j\right)\right),\label{SEdirac}\\
C_E&=&\sum_{j=1}^{\ell}\xi_j\left(1-\xi_j\right)\left(\log\frac{\xi_j}{1-\xi_j}\right)^2\label{CEdirac}.
\end{eqnarray}

The generalization to the time-dependent case, e.g., mass quenches, is straightforward. Indeed, in this case, considering the two components of the fermionic field as $\psi_k=(u_k, d_k)^T$, the corresponding components of the two-point function $C_{rs}(t)$ become \cite{Mozaffar:2021nex}
\begin{eqnarray}
&&\langle u^{\dagger}_{r}u_{s} \rangle=\frac{\delta_{rs}}{2}-\frac{1}{4\pi}\int_{-\pi}^{\pi}[\cos\theta_{0}+\sin(\theta_{0}-\theta)\sin\theta (1-\cos(2\omega_k t))]e^{i(r-s)k},\nonumber\\
&&\langle u^{\dagger}_{r}d_{s} \rangle=\frac{1}{4\pi}\int_{-\pi}^{\pi}[-\sin\theta_{0}+\sin(\theta_{0}-\theta)(\cos\theta (1-\cos(2\omega_k t))-i\sin(2\omega_k t))]e^{i(r-s)k},\nonumber\\
&&\langle d^{\dagger}_{r}u_{s} \rangle=\frac{1}{4\pi}\int_{-\pi}^{\pi}[-\sin\theta_{0}+\sin(\theta_{0}-\theta)(\cos\theta (1-\cos(2\omega_k t))+i\sin(2\omega_k t))]e^{i(r-s)k},\nonumber\\
&&\langle d^{\dagger}_{r}d_{s} \rangle=\frac{\delta_{rs}}{2}+\frac{1}{4\pi}\int_{-\pi}^{\pi}[\cos\theta_{0}+\sin(\theta_{0}-\theta)\sin\theta (1-\cos(2\omega_k t))]e^{i(r-s)k},\nonumber
\end{eqnarray}
where
\begin{equation}
\sin\theta=\frac{m^{z}}{\omega_k},\hspace*{2cm}\sin\theta_{0}=\frac{m_{0}^{z}}{\omega_{0,k}},
\end{equation}
and $\omega_{0,k}$ is given by Eq. \eqref{dispersionfermion} with $m=m_0$. 

Now we are equipped with all we need to calculate the entanglement measures both for static and time-dependent states. Unfortunately, it is not possible to find the scaling and evolution of the measures analytically in general setups. In the following we will present a combination of numerical and analytic results on the behavior of different quantities in the specific regimes of the parameter space.

\section{Capacity of entanglement for ground state} \label{sec:static}

As a first step toward understanding different aspects of capacity of entanglement in Lifshitz field theories, we would like to study this measure in static setups where the state is vacuum. We begin with the case of two-dimensional scalar theory and then, having built up some intuition, we continue by generalizing the computations to the fermionic case.

\subsection{Two-dimensional free bosons} \label{sec:bosonstatic}

In this case, the corresponding action is given by Eq. \eqref{actionscalar}. Indeed, the entanglement entropy for the ground state in this model was studied in \cite{MohammadiMozaffar:2017nri,He:2017wla}. Here, we would like to generalize these results to examine the behavior of Renyi entropy and capacity of entanglement using Eqs. \eqref{Renyi} and \eqref{CE}.

\begin{figure}[h!]
	\begin{center}
\includegraphics[scale=0.85]{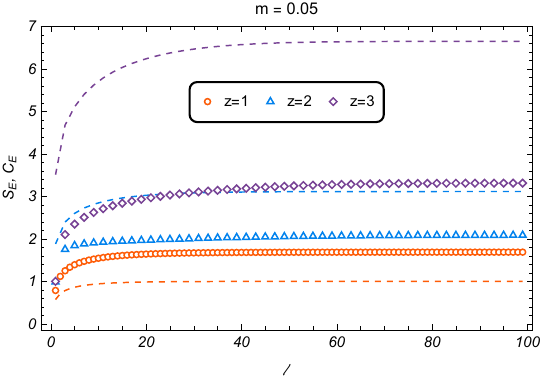}
\hspace*{0.7cm}
\includegraphics[scale=0.85]{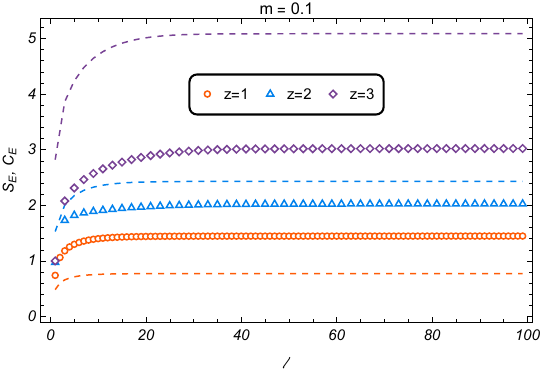}
  	\end{center}
	\caption{Capacity of entanglement as a function of $\ell$ for several values of the mass and dynamical exponent in the bosonic theory. The dashed lines represent the entanglement entropy.
	}
	\label{fig:SECEstaticbosonL}
\end{figure}

\begin{figure}[h!]
	\begin{center}
\includegraphics[scale=0.85]{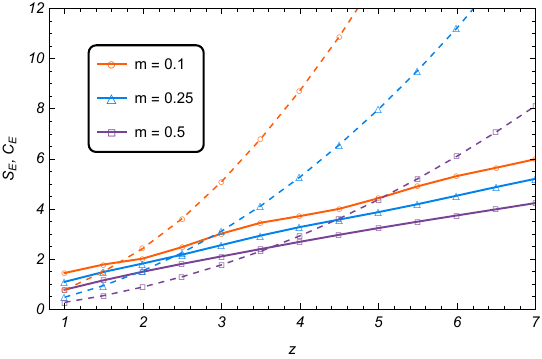}
\hspace*{0.7cm}
\includegraphics[scale=0.85]{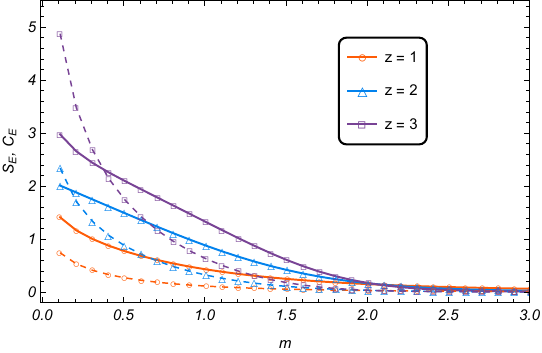}
  	\end{center}
	\caption{Left: $C_E$ as a function of $z$ for several values of the mass parameter. Right: $C_E$ as a function of mass for several values of the dynamical exponent. The dashed lines represent the entanglement entropy. Here, we set $\ell=100$.
	}
	\label{fig:SECEstaticbosonmz}
\end{figure}

We present some of the numerical results in Figs. \ref{fig:SECEstaticbosonL} and \ref{fig:SECEstaticbosonmz} for several values of the length, mass, and dynamical exponents. We will mainly consider $N=1000$, because the interesting qualitative features of the entanglement measures are independent of the total system size.\footnote{We numerically checked that
for the interested values of $\ell$, the effect of the finite $N$ is negligible.} 
Figure \ref{fig:SECEstaticbosonL} shows the entanglement entropy and capacity of entanglement
as functions of $\ell$ for several values of the dynamical exponent when the mass parameter is relatively small. We find that both of these quantities increase logarithmically in this regime. Further, the measures are monotonically increasing as we increase the dynamical exponent. Moreover, although $S_E<C_E$ for $z=1$, in the case of $z>1$, the entropy becomes much larger than the capacity. Indeed, a more careful examination shows that this behavior depends on the value of the mass parameter. 

In order to better understand the dependence of $C_E$ on the dynamical exponent, we plot this measure as a function of $z$ in the left panel of Fig. \ref{fig:SECEstaticbosonmz} for several values of the mass parameter. Note that we have also included the corresponding results for the entanglement entropy, to allow for a meaningful comparison between the different measures. Based on this figure, we see that the entanglement entropy goes on to grow indefinitely as we increase the dynamical exponent, while the capacity of entanglement grows much slower, which seems to be linear. Interestingly, for specific values of the dynamical exponent, these measures coincide, i.e., $C_E(z_c)=S_E(z_c)$, where the value of $z_c$ depends on $m$ such that for larger values of the mass parameter, it decreases. Interestingly, we see that in the large $z$ limit, $C_E\ll S_E$, which shows that the corresponding reduced density matrix becomes more and more maximally mixed as one increases the dynamical exponent.\footnote{As we have already mentioned in the Introduction section, for a maximally
mixed state the capacity of entanglement vanishes. However, here, we follow the terminology of \cite{DeBoer:2018kvc}, where the strength of the ratio between the entropy and capacity indicates how much a given density matrix is close to a maximal distribution.} Indeed, from Eq. \eqref{Hamildisscalar} one can consider $z$ as a measure for the effective correlation length (number of lattice sites that are correlated together), hence this behavior is consistent. 

The right panel of Fig. \ref{fig:SECEstaticbosonmz} illustrates entanglement entropy and capacity of entanglement as functions of mass for several values of the dynamical exponent, which shows qualitatively similar results. Moreover, both $S_E$ and $C_E$ are monotonically decreasing as we increase the mass parameter, as expected. It is worth mentioning that considering other measures, including Renyi entropy and $n$th capacity of entanglement, the qualitative features of these results do not change, although we do not explicitly show the corresponding results here.

\begin{figure}[h!]
	\begin{center}
\includegraphics[scale=0.85]{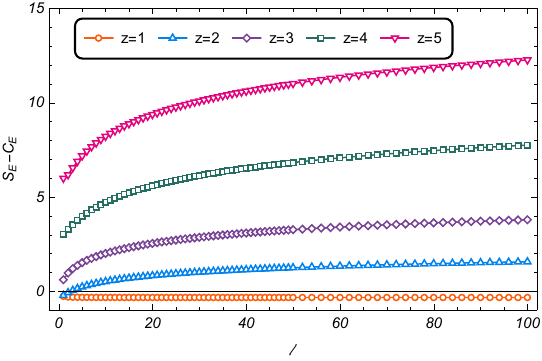}
\hspace*{0.1cm}
\includegraphics[scale=0.85]{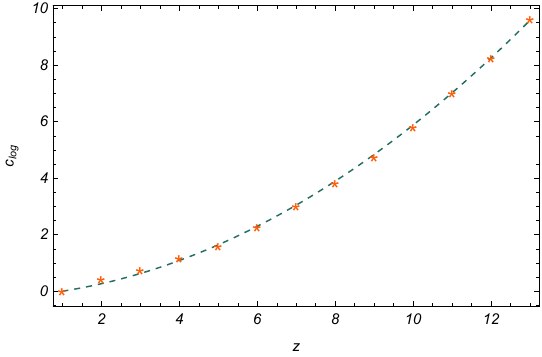}
  	\end{center}
	\caption{Left: comparing capacity of entanglement and entanglement entropy as a function of $\ell$ for several values of the dynamical exponent. Here, we consider the Dirichlet boundary
condition for the massless scalar with $N=500$.
	}
	\label{fig:SECEstaticbosonDir}
\end{figure}

To close this section, let us comment on extending this discussion to cases where the Dirichlet boundary condition is imposed on the borders of the total system. Some numerical results in the massless regime are illustrated in the left panel of Fig. \ref{fig:SECEstaticbosonDir}. Note that as we have mentioned before, in this case the translational symmetry is broken, hence the corresponding zero mode is not excited.  To facilitate comparison with the previous results, here we illustrate the difference $S_E-C_E$ as a function of $\ell$ for several values of the dynamical exponent. Again we see that for large values of the dynamical exponent, the entropy is larger than the capacity, hence the corresponding reduced density matrix becomes more maximally mixed, which is consistent with the previous results. Moreover, the difference between these measures scales logarithmically with the width of the entangling region. Indeed, as can be seen, the points are well fitted by
\begin{eqnarray}
S_E-C_E\sim c_{\rm log}\log \frac{\ell}{\epsilon}+c_0,
\end{eqnarray}
where $c_{\rm log}$ is universal and $c_0$ is scheme dependent. The right panel of Fig. \ref{fig:SECEstaticbosonDir} illustrates scaling of the universal coefficient as a function of the dynamical exponent, which is a monotonically increasing function. Indeed, based on our numerical results we see that in the relativistic limit, this universal term vanishes, hence we have $c_{\rm log}\propto (z-1)$. This observation is consistent with the previous result for a two-dimensional conformal field theory, where the capacity of entanglement tracks the entanglement entropy such that at leading order, we have $C_E=S_E$.

\subsection{Two-dimensional free fermions} \label{sec:fermionstatic}

In this section, we continue our study by computing the capacity of entanglement in a Lifshitz fermion theory Eq. \eqref{actiondirac}. To do so, we employ the corresponding expressions for the measures which are given by Eqs. \eqref{SEdirac} and \eqref{CEdirac}. It is worth mentioning that entanglement entropy in this model both for the vacuum and thermal states was studied in \cite{Vasli:2024mrf}.

\begin{figure}[h!]
	\begin{center}
\includegraphics[scale=0.85]{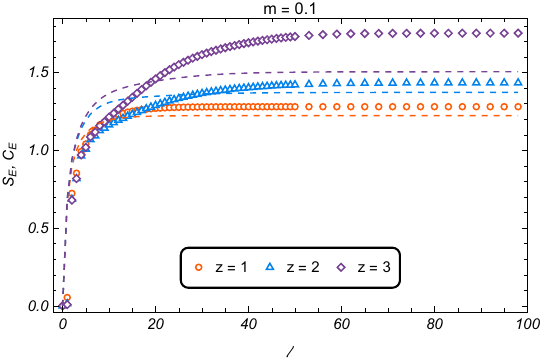}
\hspace*{0.7cm}
\includegraphics[scale=0.85]{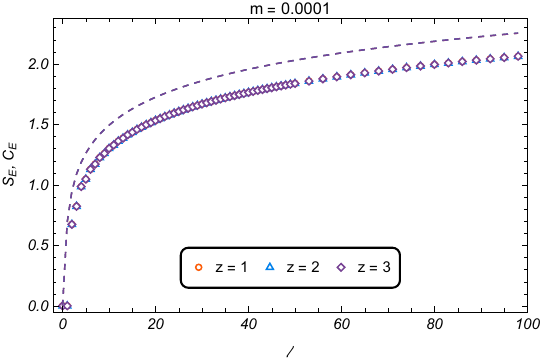}
  	\end{center}
	\caption{Capacity of entanglement as a function of $\ell$ for several values of the mass and dynamical exponent in the fermionic model. The dashed lines represent the entanglement entropy.
	}
	\label{fig:SECEstaticfermionL}
\end{figure}

In Fig. \ref{fig:SECEstaticfermionL} we summarize the corresponding numerical results for several values of the width, mass, and dynamical exponents. Note that we have also included the corresponding results for the entanglement entropy, which were previously reported in \cite{Vasli:2024mrf}, to allow for a meaningful comparison between these measures.
The left panel illustrates the capacity of entanglement as a function of $\ell$ for several values of $z$ in the massive regime. We note a number of key features: first, both these measures increase as one increases the dynamical exponent. Second, although $S_E>C_E$ for small subregions, in the case of $\ell \gg z\epsilon$, the entropy becomes smaller than the capacity. Further, this behavior becomes more pronounced as we increase the dynamical exponent. 

The right panel of Fig. \ref{fig:SECEstaticfermionL} shows the corresponding results in the small mass limit. Interestingly, the capacity of entanglement becomes completely independent of $z$. Indeed, in this case, our numerical results show that the leading behavior of $C_E$ is logarithmic, whose coefficient coincides with the relativistic case with $z=1$. Of course, similar results have been previously found for the entanglement entropy in \cite{Vasli:2024mrf}. Indeed, as shown in this reference, in the massless limit at zero temperature, the corresponding fermionic field’s two-point function is $z$-independent, thus the entanglement measures coincide with their relativistic counterparts.
Indeed, this behavior is due to the specific structure of the higher-order spatial derivatives in action \eqref{actiondirac}. Moreover, the numerical results show that in the small mass regime, the capacity becomes larger than the entropy, hence the width of the eigenvalue distribution of the corresponding reduced density matrix decreases.

\begin{figure}[h!]
	\begin{center}
\includegraphics[scale=0.85]{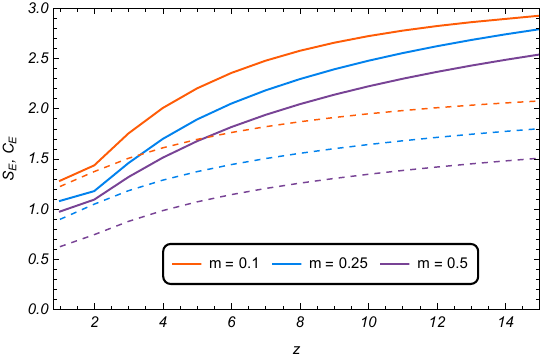}
\hspace*{0.7cm}
\includegraphics[scale=0.85]{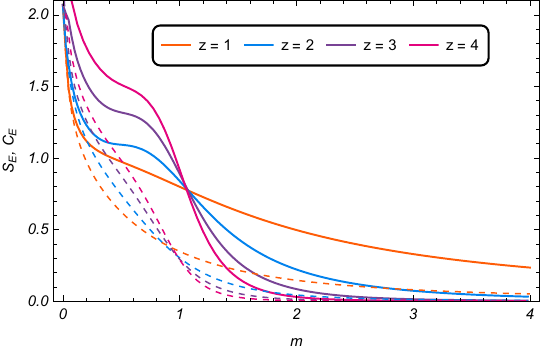}
  	\end{center}
\caption{Left: $C_E$ as a function of $z$ for several values of the mass parameter. Right: $C_E$ as a function of mass for several values of the dynamical exponent. The dashed lines represent the entanglement entropy. Here, we set $\ell=100$.
	}
	\label{fig:fermionSECEasmz1z3L100s}
\end{figure}

To better understand these behaviors, we show capacity of entanglement as a function of the dynamical exponent and the mass parameter in Fig. \ref{fig:fermionSECEasmz1z3L100s}. From the left panel, we find that this quantity increases with $z$, as expected. Although, the rate of growth is smaller than the scalar case (see the left panel in Fig.
\ref{fig:SECEstaticbosonmz}). Interestingly, we see that in the large $z$ limit, $S_E< C_E$, hence the corresponding reduced density matrix becomes more and more maximally mixed as one increases the dynamical exponent. 

The right panel of Fig. \ref{fig:fermionSECEasmz1z3L100s} illustrates the capacity of entanglement as a function of mass for several values of the dynamical exponent, which is monotonically decreasing, as expected. Again, the dashed lines represent the entanglement entropy for the same values of the parameters. Further, we see that for specific values of the mass parameter, these measures are $z$-independent. It is worth mentioning that considering other measures, including Renyi entropy and $n$th capacity of entanglement, the qualitative features of these results do not change.


\section{Entanglement $c$-functions and the nonrelativistic RG flow} \label{sec:cfunc}

In this section, we continue our study toward understanding aspects of capacity of entanglement in Lifshitz field theories by examining the existence of a consistent entropic $c$-function based on this quantity. Let us recall that in two-dimensional relativistic quantum field theories, Zamolodchikov's $c$-theorem states that there exists a function that is monotonically decreasing along the renormalization group flow such that at RG fixed points, it equals the central charge of the corresponding conformal field theory \cite{Zamolodchikov1986}. In \cite{Casini:2004bw}, the authors developed an elegant reformulation of this theorem in terms of entanglement entropy and proved an alternative entropic version of the $c$-theorem. In their construction, the $c$-function was defined as
\begin{eqnarray}\label{cfuncE}
c_S=\ell\frac{\partial S_E}{\partial \ell},
\end{eqnarray}
where, combining the Lorentz symmetry and unitarity of the underlying field theory with the strong subadditivity inequality, one can readily verify the $c$-theorem, $\partial_\ell c_S\leq 0$. Further, the above function is stationary and coincides with $\frac{c}{3}$ at the fixed point, where $c$ denotes the central charge of the corresponding conformal field theory. 

\begin{figure}[h!]
	\begin{center}
\includegraphics[scale=0.59]{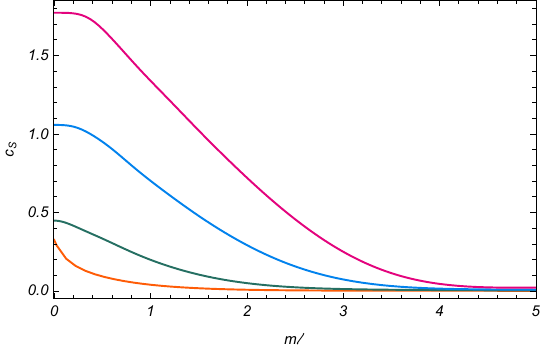}
\includegraphics[scale=0.59]{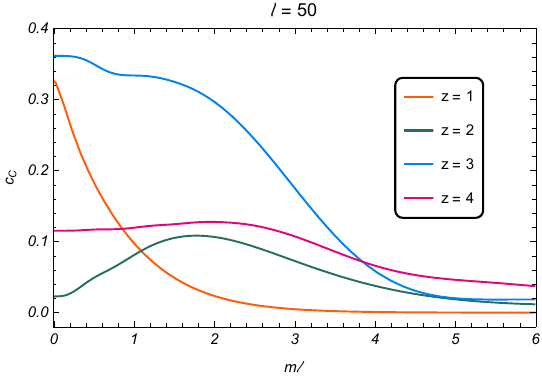}
\includegraphics[scale=0.59]{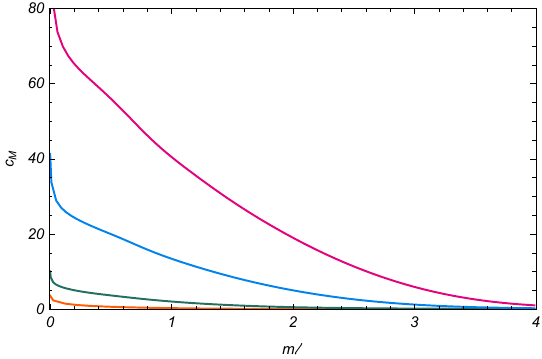}
\includegraphics[scale=0.59]{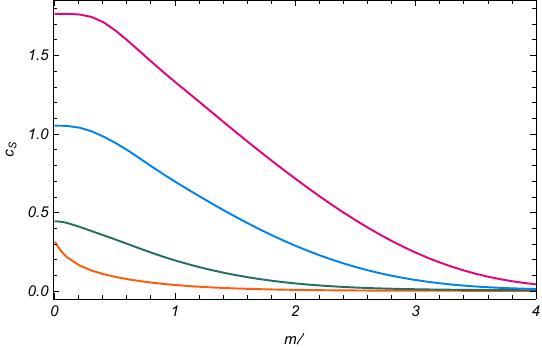}
\includegraphics[scale=0.59]{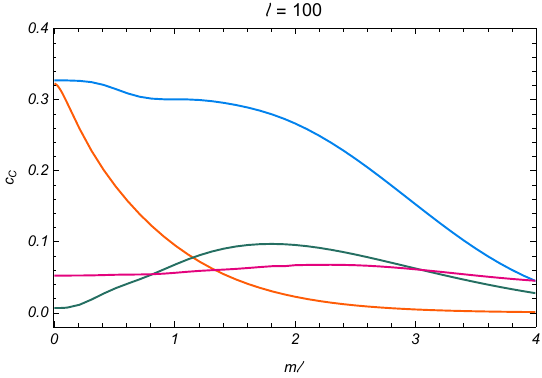}
\includegraphics[scale=0.59]{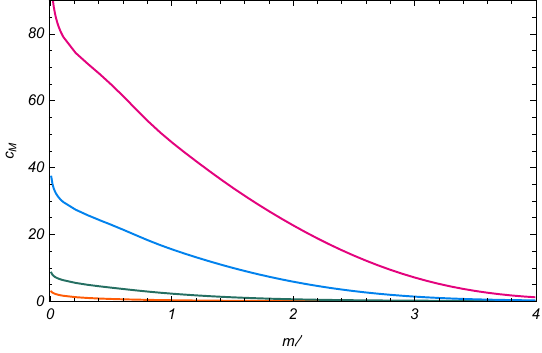}
  	\end{center}
\caption{$c_S, c_C$, and $c_M$ as functions of $m\ell$ for several values of the dynamical exponent and width of the entangling region in the bosonic theory.}
	\label{fig:bosoncfunc}
\end{figure}

\begin{figure}[h!]
	\begin{center}
\includegraphics[scale=0.43]{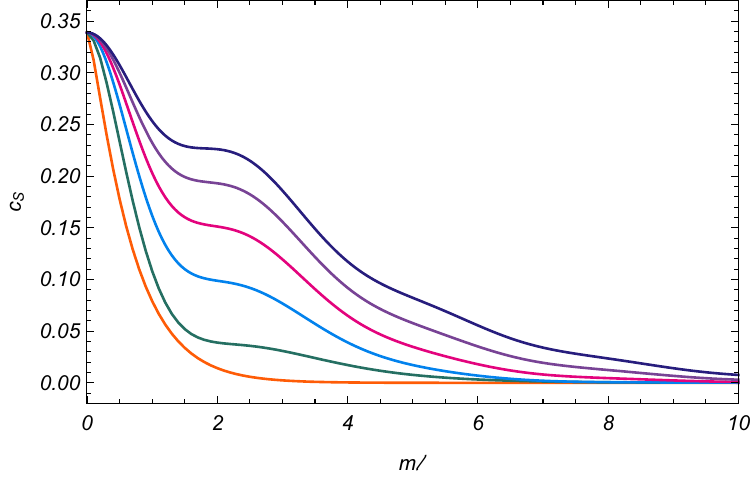}
\includegraphics[scale=0.59]{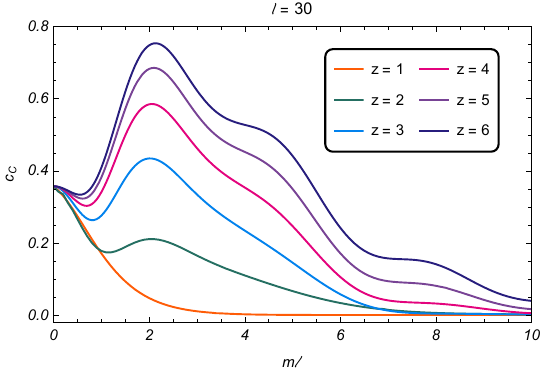}
\includegraphics[scale=0.43]{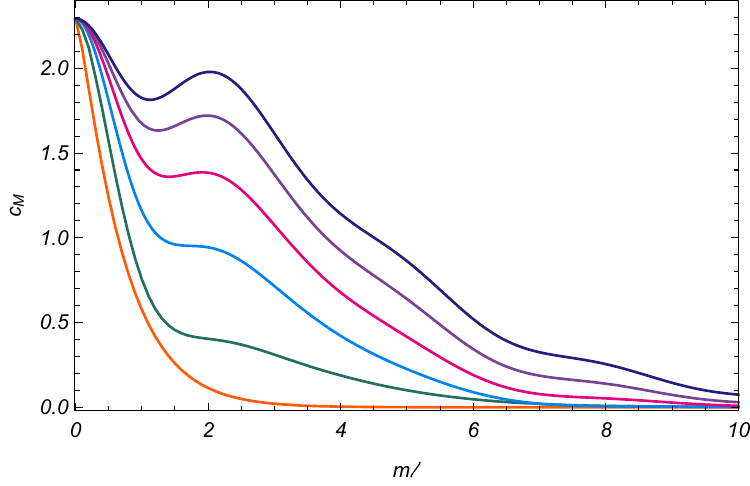}
\includegraphics[scale=0.59]{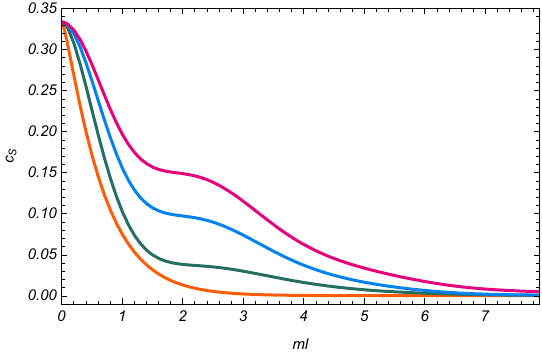}
\includegraphics[scale=0.59]{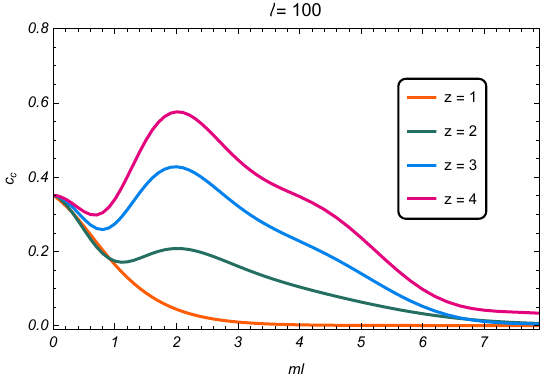}
\includegraphics[scale=0.59]{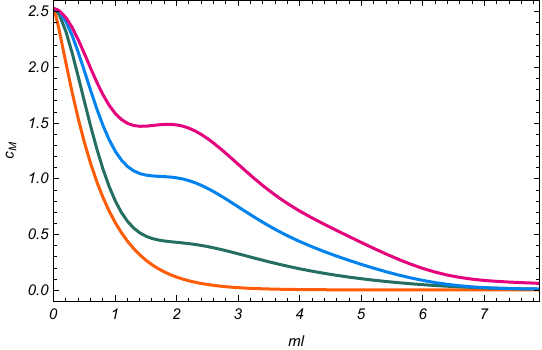}
  	\end{center}
\caption{$c_S, c_C$, and $c_M$ as functions of $m\ell$ for several values of the dynamical exponent and width of the entangling region in fermionic theory.}
	\label{fig:fermioncfunc}
\end{figure}

This line of thought is useful for providing other entanglement-based proof for the $c$-theorem and its higher-dimensional counterparts, including $F$- and $a$-theorems \cite{Casini:2012ei,Casini:2017vbe}. Remarkably, we can construct entanglement candidate $c$-functions using other entanglement measures, e.g., mutual information \cite{Casini:2015woa}.\footnote{ Let us mention that for relativistic theories that are not scale invariant, another $c$-function based on
entanglement measures was also defined in \cite{Castro-Alvaredo:2011wqh}. This definition is connected to the definition
of entanglement measures in massive quantum field theory by employing
the branch point twist field approach introduced in \cite{Cardy:2007mb}.} In particular, in \cite{Arias:2023kni}, the authors define a $c$-function based on
the capacity of entanglement similar to the one based on entropy, which displays a monotonic behavior under the RG flow generated by the mass parameter. In this case, the corresponding $c$-function is defined as follows:
\begin{eqnarray}\label{cfuncC}
c_C=\ell\frac{\partial C_E}{\partial \ell},
\end{eqnarray}
which is monotonically decreasing along the RG flow and at the fixed point reduces to the central charge of the theory. Indeed, in a two-dimensional conformal field theory with $C_E=S_E$ at leading order, this behavior is not surprising. It is worth mentioning that, unlike the entanglement entropy, the capacity of entanglement does not satisfy any of the concavity properties, hence its monotonic behavior seems to be accidental. Further, in \cite{Arias:2023kni}, another entanglement-based $c$-function has been proposed, which is the second moment of the shifted modular Hamiltonian, defined as follows:
\begin{eqnarray}\label{cfuncC}
M_E=\langle(1+H_A)^2\rangle=(S_E+1)^2+C_E,
\end{eqnarray}
where in the second equality, we used Eqs. \eqref{EE} and \eqref{capa}. In this case, the corresponding $c$-function is 
\begin{eqnarray}\label{cfuncM}
c_M=\ell\frac{\partial M_E}{\partial \ell}=2(1+S_E)c_S+c_C.
\end{eqnarray}
Note that based on the above definition, $c_M$ depends explicitly on the entanglement entropy and, as a consequence, it diverges in the continuum limit. A detailed investigation exhibits that the above quantity satisfies monotonicity in specific theories. It seems that this property is due to the Schur concavity of $M_E$ \cite{Boes}, although it does not respect the strong subadditivity necessarily. Here, we would like to apply the approaches developed in these references to examine the behavior of entanglement $c$-functions in the presence of a nontrivial dynamical exponent. We present some of the numerical results for the $c$-functions in Figs. \ref{fig:bosoncfunc} and \ref{fig:fermioncfunc}, respectively.

Figure \ref{fig:bosoncfunc} illustrates the corresponding results in the bosonic theory for several values of the dynamical exponent and width of the entangling region. The plots in the first row correspond to $\ell=50$. The left panel shows $c_S$, which displays a monotonic behavior under the RG flow both for relativistic and nonrelativistic scaling. Indeed, in this case we see that $c_S$ increases with $z$ such that for each value of the dynamical exponent, the model is flowing to a massive theory in the IR limit with a smaller number of degrees of freedom. Note that this behavior is perfectly consistent with the previous results reported in the literature, e.g., \cite{Arias:2023kni} and \cite{Vasli:2024mrf} for $z=1$ and $z>1$, respectively. Let us mention that in the nonrelativistic case, the strong subadditivity of the entanglement entropy does not necessarily require $c_S$ to be a convenient $c$-function. 

Turning to the $c$-function based on the capacity of entanglement, the middle panel in Fig. \ref{fig:bosoncfunc} presents $c_C$ for the same values of the parameters. Interestingly, although this function decreases monotonically for the case of $z=1$, for larger values of the dynamical exponent it fails to satisfy this property. Hence, we may conclude that for $z>1$, where the Lorentz invariance is expected to break down, $c_C$ is not a genuine $c$-function. Also, a more careful examination shows that this quantity is not a monotonically increasing function of $z$. Finally, the right panel shows that $c_M$ displays a monotonic behavior under the RG flow generated by the mass parameter. Indeed, as
can be seen from Eq. \eqref{cfuncM}, the contribution to $c_M$ due to $c_S$ has a nontrivial weight, which depends on the entanglement entropy. Hence, with a large enough $S_E$, $c_M$ behaves in the same manner as $c_S$ along the RG flow. 
Further, the plots in the second row correspond to $\ell=100$, which shows qualitatively similar behaviors.

Figure \ref{fig:fermioncfunc} illustrates the corresponding results in the fermionic theory for several values of the dynamical exponent and width of the entangling region.\footnote{The results for $c_S$ with $\ell=30$ were previously reported in \cite{Vasli:2024mrf}.} The first row corresponds to $\ell=30$. The left panel illustrates $c_S$, which displays a monotonic behavior under the RG flow in the relativistic case with $z=1$. Similar behavior for $c_S$ is found in the nonrelativistic case, except that it has a plateau at some specific value of the mass parameter. Remarkably, we see that the location of the plateau regime is independent of $z$. Again, in this case, we see that $c_S$ increases with $z$ such that for each value of the dynamical exponent, the theory is flowing to a massive theory in the IR limit with a smaller number of degrees of freedom. 

The middle panel in Fig. \ref{fig:fermioncfunc} presents $c_C$ for the same values of the parameters. The key feature of our numerical results is that although $c_C'<0$ for $z=1$, this is no longer the case for larger values of the dynamical exponent. Once again, we see that for $z>1$, $c_C$ is not a candidate $c$-function. Moreover, the right panel shows that $c_M$ tracks $c_S$ such that it displays a monotonic behavior under the RG flow and presents a plateau behavior at some specific value of the mass parameter. Indeed, for $\ell=30$, the entanglement entropy is not large enough, and the second term in Eq. \eqref{cfuncM} is dominant. Thus, the main contribution to $c_M$ is due to $c_C$, which gives a similar dependence on the mass parameter. Further, the plots in the second row correspond to $\ell=100$, which shows that the qualitative behavior of $c$-functions does not depend strongly on $\ell$.

\section{Capacity of entanglement after a global quantum quench} \label{sec:quench}

In this section, we study the dynamics of capacity of entanglement followed by a global quantum quench in bosonic and fermionic models. Moreover, to get a better understanding of the results, we will also compare
the behavior of different measures, e.g., entanglement entropy, in various setups. To do so, we provide a numerical analysis and examine the various regimes in the growth of capacity of entanglement using the quasiparticle picture. 

\subsection{Entanglement dynamics and quasiparticle picture}\label{sec:quench1}

Let us recall that the evolution of entanglement measures after a global quench can be understood in terms of the quasiparticles, interpretation for the propagation of entanglement \cite{Calabrese:2005in}. Indeed, for a wide variety of global quenches, this picture provides an understanding of the main qualitative features of the entanglement dynamics. Further, this picture has been improved in \cite{Alba:2017ekd}, where it has been suggested that for certain integrable models, the distribution of quasiparticles can be conjectured from the equivalence between the entanglement and the
thermodynamic entropy in the late-time stationary state, which is described by a generalized Gibbs ensemble.
Based on this remarkable idea for a connected interval of width $\ell$ embedded in an infinite line, the time-dependent entanglement entropy is given by
\begin{eqnarray}\label{SEt}
S_E(t)=2t\int_{2|v(k)|t<\ell}s(k)v(k)\,dk+\ell\int_{2|v(k)|t>\ell}s(k)\,dk,
\end{eqnarray}
where $v(k)\equiv \frac{d\omega_k}{dk}$ is the group velocity of the quasiparticles. Moreover, $s(k)$ is the entropy density, which depends on the production rate of quasiparticles and can be evaluated explicitly from the mode occupation number (the expectation value of the number operator in the initial state), yielding \cite{Alba:2017lvc}
\begin{eqnarray}\label{sk}
2\pi s(k)=-n_k\ln n_k\pm (1\pm n_k)\ln (1\pm n_k),
\end{eqnarray}
where the upper and the lower signs correspond to the bosonic and the fermionic cases, respectively.
Explicit expressions for the mode occupation numbers read
\begin{eqnarray}\label{nk}
n_k^{(B)}=\frac{1}{4}\left(\frac{\omega_k}{\omega_{0,k}}+\frac{\omega_{0,k}}{\omega_k}\right)-\frac{1}{2},\hspace*{1cm}n_k^{(F)}=\frac{1}{2}-\frac{1}{2}\frac{\omega_k}{\omega_{0,k}}.
\end{eqnarray}
Also, the corresponding group velocity is easily obtained by differentiating the dispersion relations with respect to $k$, which gives
\begin{eqnarray}\label{vg}
v(k)^{(B)}=\frac{z\left(2\sin \frac{k}{2}\right)^{2z-2}\sin k}{\sqrt{m^{2z}+\left(2\sin \frac{k}{2}\right)^{2z}}},\hspace*{1cm}v(k)^{(F)}=\frac{z\left(\sin k\right)^{2z-1}\cos k}{\sqrt{m^{2z}+\left(\sin k\right)^{2z}}}.
\end{eqnarray}
Combining the above results with Eqs. \eqref{SEt} and \eqref{sk}, one can find an analytic description for the time evolution
of entanglement entropy and other related quantities, including the mutual information \cite{Alba:2019ybw}. Also, this approach may help us to study entanglement revivals, which provides a useful tool to diagnose quantum information scrambling in quantum many-body systems \cite{Modak:2020faf}. Interestingly, it was recently found that precisely the same approach can be applied to obtain a similar expression for the evolution of capacity of entanglement after a global quantum quench, as follows \cite{Arias:2023kni}:
\begin{eqnarray}\label{CEt}
C_E(t)=2t\int_{2|v(k)|t<\ell}c(k)v(k)\,dk+\ell\int_{2|v(k)|t>\ell}c(k)\,dk,
\end{eqnarray}
where $c(k)$ denotes the capacity density, whose explicit expression is given by
\begin{eqnarray}\label{ck}
2\pi c(k)=n_k(1\pm n_k)\left(\ln\frac{1\pm n_k}{n_k}\right)^2.
\end{eqnarray}
An interesting observation in \cite{Arias:2023kni} was that although both $S_E(t)$ and $C_E(t)$ exhibit an intermediate linear growth, the slopes characterizing the linear regime are different and, consequently, also the late-time saturation value. Thus, the conformal field theory prediction for critical evolutions is not confirmed by the numerical results.

\begin{figure}[h!]
	\begin{center}
\includegraphics[scale=0.58]{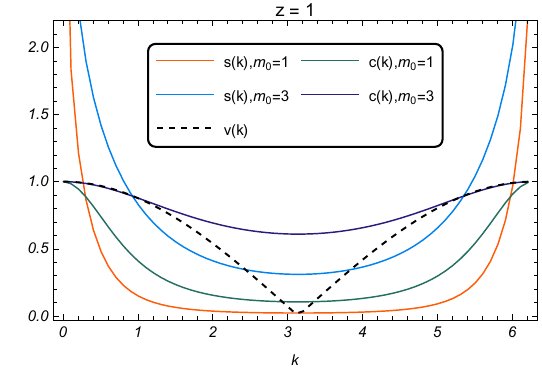}
\includegraphics[scale=0.58]{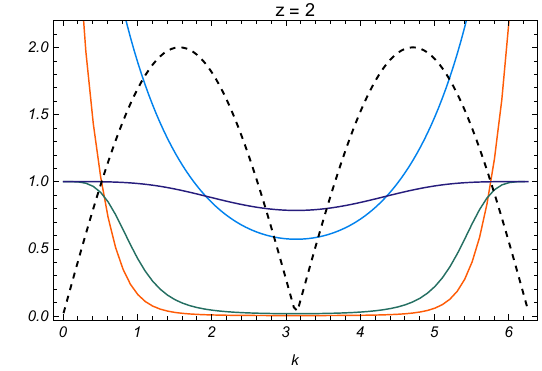}
\includegraphics[scale=0.58]{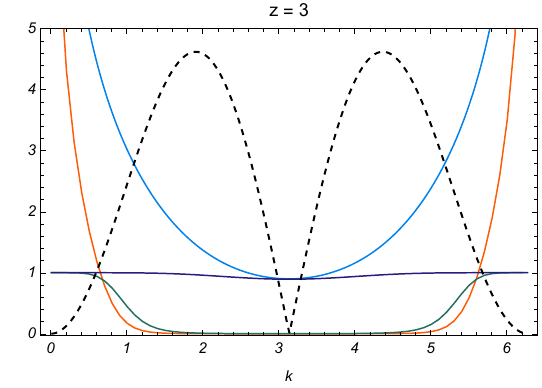}
  	\end{center}
\caption{Entropy and capacity densities as functions of $k$ for different values of $z$ and $m_0$ in bosonic theory. The dashed line indicates the group velocity. Here, we consider the critical case with $m=0$.
	}
	\label{fig:SCtildeboson}
\end{figure}

\begin{figure}[h!]
	\begin{center}
\includegraphics[scale=0.58]{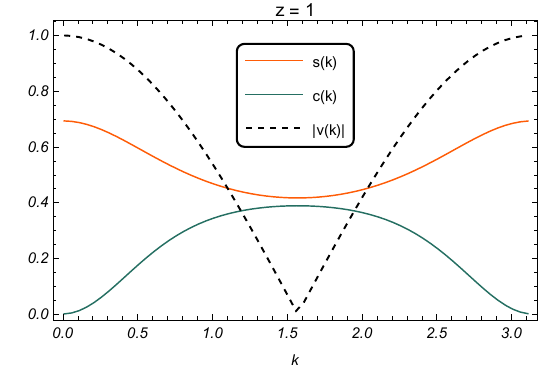}
\includegraphics[scale=0.58]{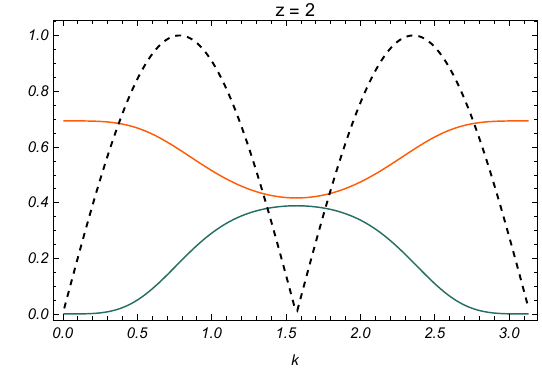}
\includegraphics[scale=0.58]{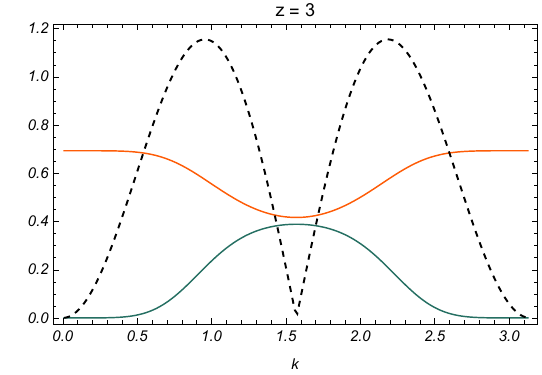}
  	\end{center}
\caption{Entropy and capacity densities as functions of $k$ for different values of $z$ in fermionic theory. The dashed line indicates the group velocity. Here, we consider the critical case with $m=0$ and $m_0=1$.
	}
	\label{fig:SCtildefermion}
\end{figure}

\begin{figure}[h!]
	\begin{center}
\includegraphics[scale=0.58]{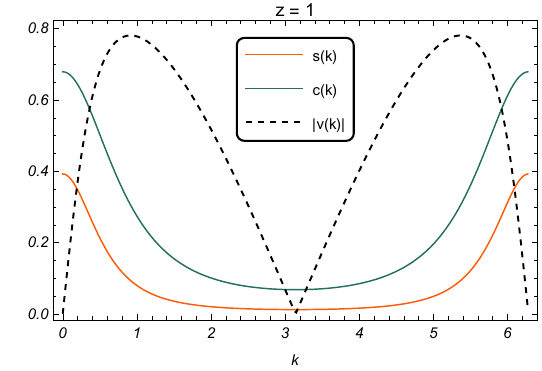}
\includegraphics[scale=0.58]{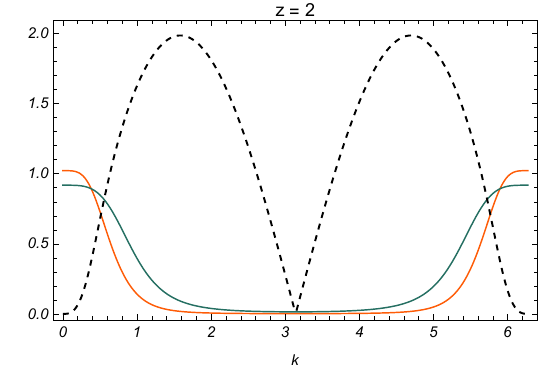}
\includegraphics[scale=0.58]{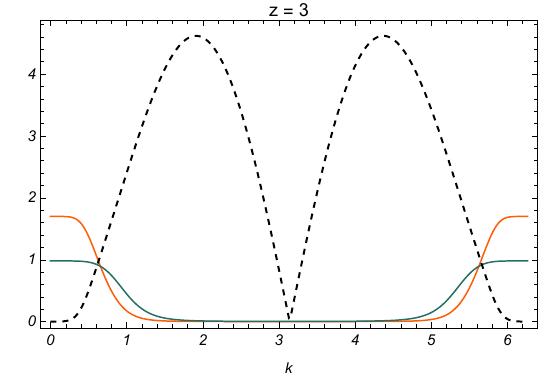}
  	\end{center}
\caption{Entropy and capacity densities as functions of $k$ for different values of the dynamical exponent in a bosonic theory when the postquench Hamiltonian is massive. The dashed line indicates the group velocity. Here, we set $m=0.5$ and $m_0=1$.
	}
	\label{fig:SCtildebosonmassive}
\end{figure}

It is straightforward to generalize the above approach to study temporal evolution of different entanglement measures after a global quench in Lifshitz theories, which has been previously done for the entanglement entropy in \cite{MohammadiMozaffar:2018vmk,Mozaffar:2021nex}. Indeed, in these references it was shown that due to a nontrivial dynamical exponent, the propagation of quasiparticles is completely different from the relativistic case with $z=1$. Hence, the corresponding entanglement entropy for Lifshitz models exhibits an extremely slower late-time saturation extending over infinite time due to the presence of a pile of slow modes with vanishingly small velocity. In the following, we will investigate how these specific properties of the spectrum of the quasiparticles will modify the time scaling of the capacity of entanglement. Before examining the full time dependence of the measures, we would like to study in more detail the $z$ dependence of the corresponding quantities. The numerical results for different quantities are summarized in Figs. \ref{fig:SCtildeboson}, \ref{fig:SCtildefermion}, and \ref{fig:SCtildebosonmassive}. 

Figure \ref{fig:SCtildeboson} illustrates the behavior of the entropy and capacity density as functions of momentum for several values of $z$ and $m_0$ with $m=0$. Also, the dashed line indicates the group velocity for the same values of the parameters. In this case, when the postquench Hamiltonian is scale invariant, we find a number of key features. First, although the entropy density diverges in $k=0$ and $k=2\pi$ limits, the capacity density saturates to unity (where it reaches its maximum value). Indeed, a perturbative expansion for small $k$ yields
\begin{eqnarray}\label{skckexpand}
s(k\sim 0)\sim -z\log\frac{k}{m_0}+\cdots, \hspace*{2cm}c(k\sim 0)\sim 1-\frac{4}{3}\left(\frac{k}{m_0}\right)^{2z}+\cdots,
\end{eqnarray}
which shows that the production rate of quasiparticles and their individual contributions to $s(k)$ and $c(k)$ are completely different. In particular, the subleading term in the expansion of capacity density shows that its maximum value is approached from below, and that this behavior is controlled by the dynamical exponent. Second, for generic integer values of $z$, the region with maximum velocity does not necessarily coincide with the region where entropy and capacity densities reach their maximum. In order to better understand this behavior, we use the expansion for group velocity of the quasiparticles near $k\rightarrow 0$, which reads 
\begin{eqnarray}\label{vkexpand}
v(k\sim 0)\sim zk^{z-1}-\frac{z(z+2)}{24}k^{z+1}+\cdots.
\end{eqnarray}
Indeed, comparing the above expression with Eq. \eqref{skckexpand}, we see that only in the relativistic case does the region with maximum velocity ($v\sim 1$) coincide with the region where entropy and capacity densities reach their maximum (see also the left panel in Fig. \ref{fig:SCtildeboson}). Thus, for $z=1$ the contribution of fast quasiparticles to $S_E(t)$ and $C_E(t)$ is dominant, and the entanglement is mainly carried by these modes. On the other hand, for $z>1$ the corresponding quasiparticles reach their maximum velocity at $k_{\rm max}=\arccos \frac{2-z}{z}$ \cite{MohammadiMozaffar:2018vmk}. Using this result, one can show that the group velocity peaks at $v_{\rm max}=\sqrt{z}\left(4\frac{z-1}{z}\right)^{\frac{z-1}{2}}$,
which coincides with numerical results depicted in the middle and right panels in Fig. \ref{fig:SCtildeboson}. Also, the width of the peak is controlled by the dynamical exponent. From these plots, we see that at $k=0$, where $s(k)$ and $c(k)$ reach their maximum values, the velocity vanishes, which is consistent with Eq. \eqref{vkexpand}. Based on this observation, we expect that in this nonrelativistic setup, most of the entanglement is carried by the slow modes \cite{Mozaffar:2021nex}. Further, both the entropy and capacity densities are monotonically increasing functions of the dynamical exponent and prequench mass, as expected. 

In a similar manner, one can extend this analysis to the fermionic theories. In this case, the behavior of $s(k)$, $c(k)$, and the group velocity for several values of the dynamical exponent are summarized in Fig. \ref{fig:SCtildefermion}. A key observation to note here is that while the entropy density remains finite at $k\sim \{0, \pi\}$, the capacity density vanishes in these limits. It is instructive to expand these functions in the small momentum limit, where one obtains
\begin{eqnarray}\label{skckexpandfer}
s(k\sim 0)\sim \log 2-\frac{1}{2}\left(\frac{k}{m_0}\right)^{2z}+\cdots, \hspace*{2cm}c(k\sim 0)\sim \left(\frac{k}{m_0}\right)^{2z}+\cdots.
\end{eqnarray}
Moreover, we expand the velocity of quasiparticles around $k\sim 0$ up to second order as
\begin{eqnarray}\label{vkexpandfer}
v(k\sim 0)\sim zk^{z-1}-\frac{z(z+2)}{6}k^{z+1}+\cdots.
\end{eqnarray}
Interestingly, although for $z=1$ the contribution of fast modes to $S_E(t)$ is dominant, their contribution to $C_E(t)$ is much less pronounced, hence this quantity is mainly carried by the slow modes near $k\sim \frac{\pi}{2}$ (see the left panel in Fig. \ref{fig:SCtildefermion}). On the other hand, from the middle and right panels in this figure, we see that for larger values of the dynamical exponent, the contribution of fast modes to the capacity of entanglement becomes more pronounced. It is relatively simple to show that in this case, the corresponding quasiparticles reach their maximum velocity at $k_{\rm max}=\frac{1}{2}\arccos \frac{2-z}{z}$, thus we have $v_{\rm max}=\sqrt{z}\left(\frac{z-1}{z}\right)^{\frac{z-1}{2}}$. Hence, similar to the bosonic case for $z>1$, the contribution of slow modes to the entanglement entropy is dominant. 

We can also extend this analysis to other cases where the postquench mass is finite. For example, in Fig. \ref{fig:SCtildebosonmassive} we report the behavior of the densities and velocity for a quench from $m_0=1$ to $m=0.5$ in the bosonic theory. Comparing to the critical case, a crucial difference is that even for the relativistic case, the region with
maximum densities coincides with the region where the velocity vanishes, hence most of the entanglement is carried by the slow modes. This behavior becomes more pronounced as we increase the dynamical exponent. Further, similar to the previous case, the group velocity is an increasing function of $z$. It is worthwhile to mention that the maximum velocity decreases upon increasing the postquench mass parameter. Indeed, a perturbative expansion around $m \sim 0$ yields \cite{MohammadiMozaffar:2018vmk}
\begin{eqnarray}\label{vkmax}
v_{\rm max}=\sqrt{z}\left(4\frac{z-1}{z}\right)^{\frac{z-1}{2}}-\frac{\sqrt{z}}{2^{z+2}}\left(\frac{z-1}{z}\right)^{-\frac{z+1}{2}}m^{2z}+\cdots,
\end{eqnarray}
which shows that the small mass corrections rapidly decrease upon increasing $z$. Let us emphasize that we found similar results for the fermionic theory, although we do not explicitly show the corresponding figures here. 

\subsection{Dynamics of the capacity of entanglement}\label{sec:quench2}

\begin{figure}[h!]
	\begin{center}
\includegraphics[scale=0.85]{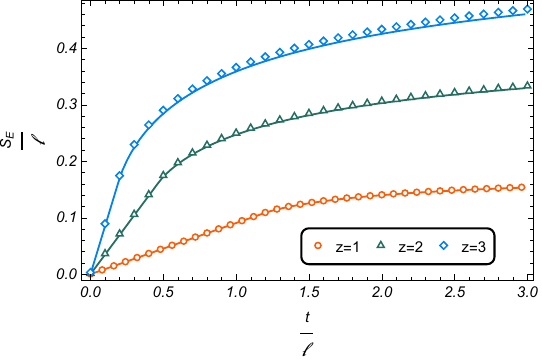}
\hspace*{0.7cm}
\includegraphics[scale=0.85]{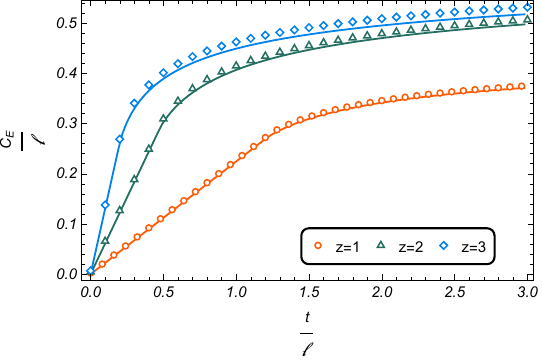}
  	\end{center}
\caption{Temporal evolution of entanglement entropy (left) and capacity of entanglement (right) for several values of the dynamical exponent in scalar theory with $\ell=200$. The solid lines are obtained from the quasiparticle picture. Here, we set $m=2m_0=2$.
	}
	\label{fig:SECEtimel200z123}
\end{figure}

\begin{figure}[h!]
	\begin{center}
\includegraphics[scale=0.58]{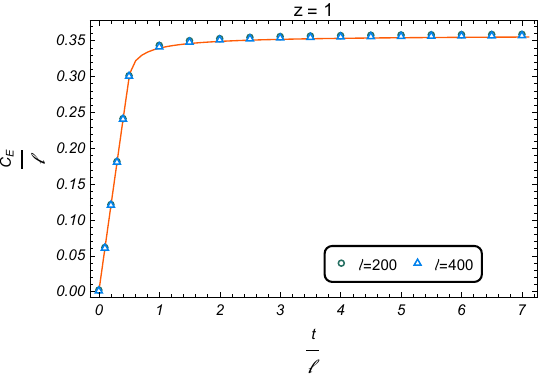}
\includegraphics[scale=0.58]{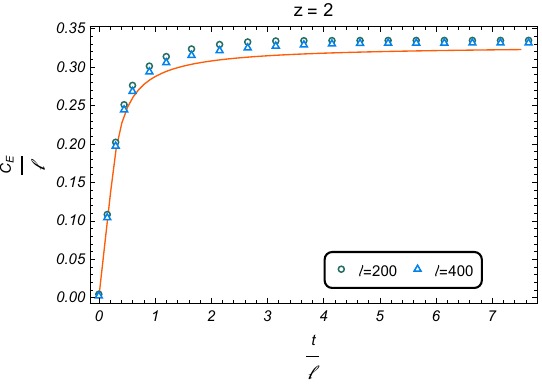}
\includegraphics[scale=0.58]{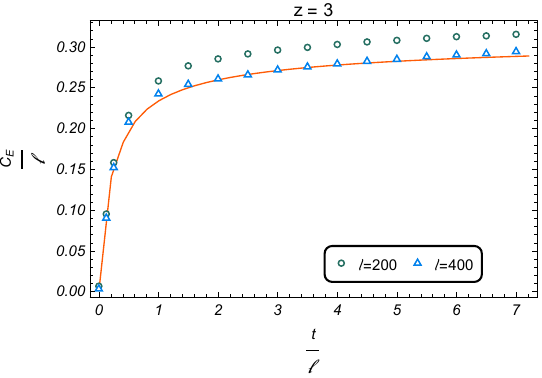}
\includegraphics[scale=0.58]{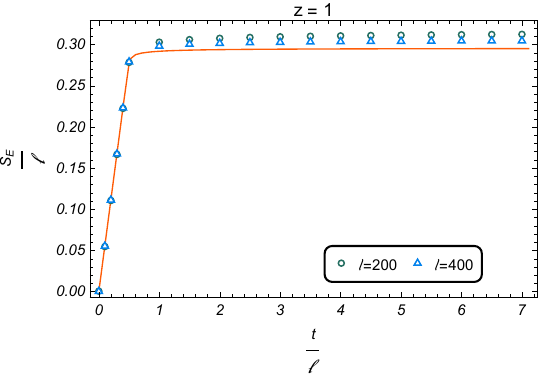}
\includegraphics[scale=0.58]{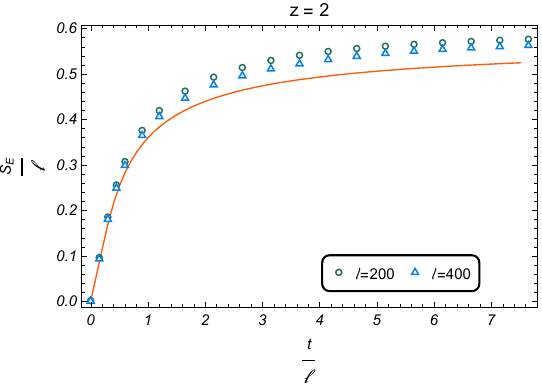}
\includegraphics[scale=0.58]{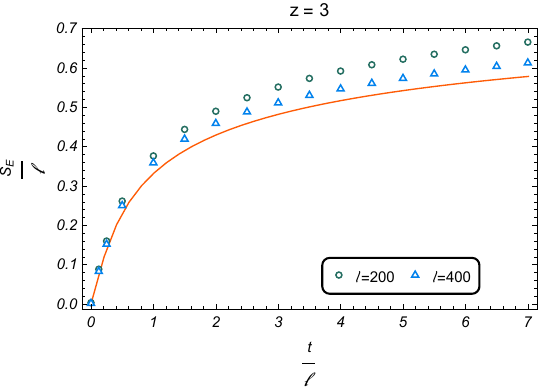}
  	\end{center}
\caption{Temporal evolution of capacity of entanglement and entanglement entropy for several values of $z$ and $\ell$ in scalar theory. The solid lines are obtained from the quasiparticle picture. Here, we set $m_0=1$ and $m=10^{-6}$.
	}
	\label{fig:SECEtimemasslessz13}
\end{figure}

Let us now turn to the computation of $C_E(t)$ in this setup using the correlator method. The corresponding numerical results for the bosonic theory are summarized in Figs. \ref{fig:SECEtimel200z123} and \ref{fig:SECEtimemasslessz13} for the massive and critical quenches, respectively. Note that we have also included the corresponding results for the
evolution of entanglement entropy, which was previously reported in \cite{Mozaffar:2021nex,MohammadiMozaffar:2018vmk}, to allow for a meaningful comparison between the different measures. 

Figure \ref{fig:SECEtimel200z123} illustrates both the numerical results and the quasiparticle prediction for the entanglement dynamics after the global quench when the postquench Hamiltonian is massive. The figure shows the densities
$S_E(t)/\ell$ and $C_E(t)/\ell$ plotted versus the rescaled time $t/\ell$. The markers correspond to numerical results and the solid lines have been obtained from Eqs. \eqref{SEt} and \eqref{CEt}. The different curves in the figure correspond to quenches with different values of the dynamical exponent. We observe that the time evolution of capacity and entropy is characterized by two different scaling regimes: an early-time linear growth and a late-time saturation. The crossover time from the linear to the saturation regime decreases with $z$, because the maximum velocity $v_{\rm max}$ increases upon increasing the dynamical exponent, as it is clear from Eq. \eqref{vkmax}. Of course, the values of both measures increase with the dynamical exponent as expected. Indeed, the entanglement entropy is more sensitive to the variation of $z$, which is clear from the time scaling during the saturation regime. Further, let us add that for any finite $\ell$, corrections are expected because the quasiparticle predictions Eqs. \eqref{SEt} and \eqref{CEt} hold only in the regime where $t, \ell \rightarrow\infty$ with $t/\ell$ fixed. The small departure from the numerical results is due to these corrections, which is rapidly decrease upon increasing the width of the entangling region.

To gain more insights into the evolution of the measures, let us turn our attention to the critical quenches, where the postquench Hamiltonian is massless. The corresponding results are summarized in Fig. \ref{fig:SECEtimemasslessz13}.\footnote{Note that for the critical quench when the postquench Hamiltonian is scale invariant, logarithmic corrections due to the existence of zero modes appear \cite{Cotler:2016acd,MohammadiMozaffar:2018vmk}. Hence, we will stay away from the critical case in the numerical analysis.} Although in the case of $z=1$ the saturation happens almost instantaneously, for larger values of the dynamical exponent, lattice simulation shows a mild transition. As the previous quasiparticle analysis suggested, one interesting difference for $z>1$, with respect to the relativistic case, is that the contribution of slow modes to the spread of entanglement is dominant. As a result, we expect a smooth transition between the linear growth and the saturation regimes. Indeed, this behavior is consistent with the previous results for $S_E(t)$ reported in \cite{Mozaffar:2021nex}, where the transition between these regimes depends on the velocity distribution of the quasiparticles, such that in specific cases with large values of the dynamical exponent, we have an infinite saturation time. Of course, a similar scaling behavior has been previously found for $C_E(t)$ in the relativistic case \cite{Arias:2023kni}. 

\begin{figure}[h!]
	\begin{center}
\includegraphics[scale=0.58]{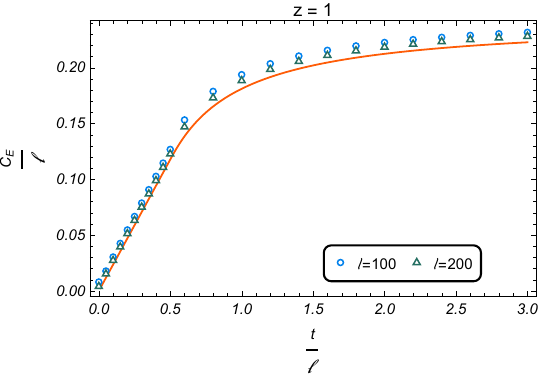}
\includegraphics[scale=0.58]{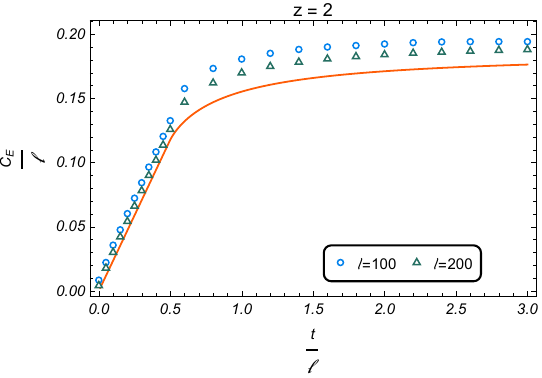}
\includegraphics[scale=0.58]{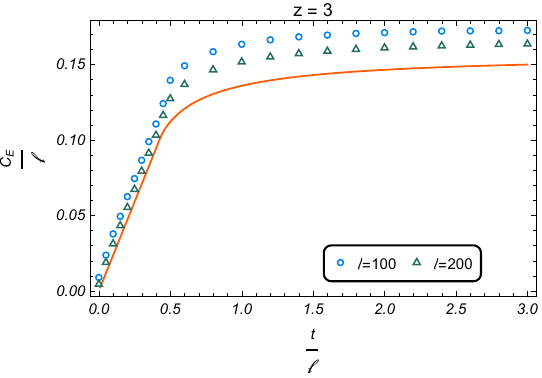}
\includegraphics[scale=0.58]{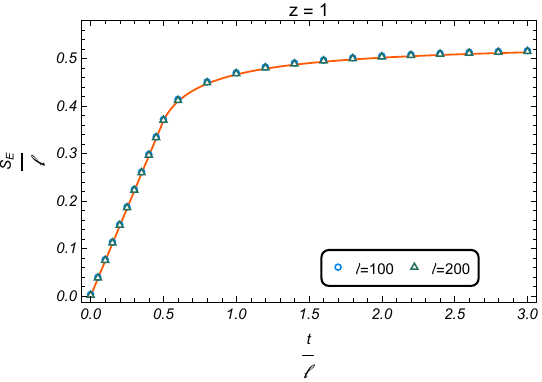}
\includegraphics[scale=0.58]{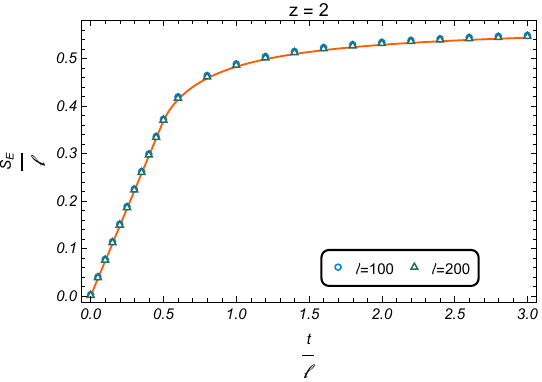}
\includegraphics[scale=0.58]{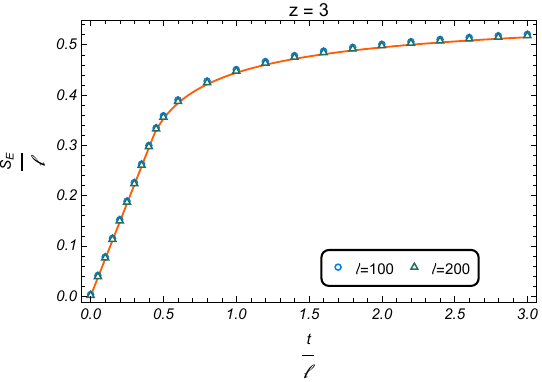}
  	\end{center}
\caption{Temporal evolution of capacity of entanglement and entanglement entropy for several values of $z$ and $\ell$ in Dirac Lifshitz theory. The solid lines are obtained from the quasiparticle picture. Here, we set $m_0=1$ and $m=0$.
	}
	\label{fig:fermionSECEtimemasslessz13}
\end{figure}

In a similar manner, one can extend this analysis to the fermionic theories. The corresponding numerical results are summarized in Fig. \ref{fig:fermionSECEtimemasslessz13} for several values of the parameters when the quench is critical.\footnote{Let us note that in fermionic theories when the postquench mass is greater that the initial mass, Eq. \eqref{nk} yields a negative occupation number, hence we are assuming $m<m_0$ here.} The markers in this figure show the numerical results, which in the large $\ell$ limit coincide with the quasiparticle picture given by Eqs. \eqref{SEt} and \eqref{CEt} (represented by the continuous orange line). Again, we see a mild transition between the two scaling regimes. One can see that in the period of linear growth, the slope seems more or less the same independent of the width of the entangling region. Moreover, the slopes of the linear growth for the entropy and capacity are different, hence so is the saturation value. Interestingly, during the saturation regime the rate of change of the capacity decreases with the dynamical exponent. Of course, this is in agreement with the results shown in Fig. \ref{fig:SCtildefermion}, where we see that for larger values of the dynamical exponent, the contribution of fast modes to $C_E$ becomes more pronounced. The situation is completely different from the bosonic case, where most of the entanglement is carried by the slow modes (see Fig. \ref{fig:SCtildeboson}).

\begin{figure}[h!]
	\begin{center}
\includegraphics[scale=0.78]{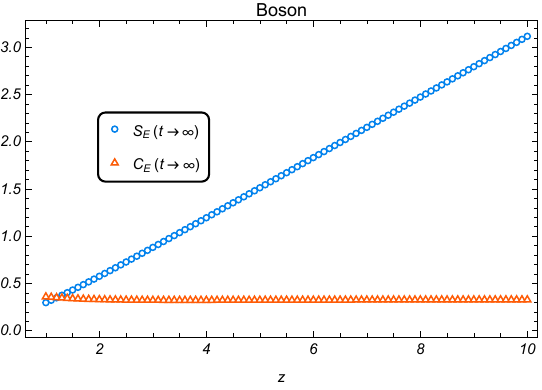}
\hspace*{1cm}
\includegraphics[scale=0.78]{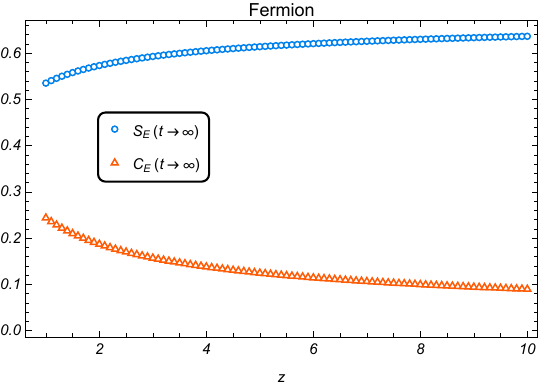}
  	\end{center}
\caption{Saturation constants of $S_E$ and $C_E$ as functions of the dynamical exponent, obtained by fixing the parameters in Eqs. \eqref{SEt} and \eqref{CEt}. Here, we set $m_0=1$ and $m=10^{-6}$.
	}
	\label{fig:saturation}
\end{figure}

To get a better understanding of the late-time behavior, we compare the values of saturation constants of $S_E$ and $C_E$ as functions of the dynamical exponent in Fig. \ref{fig:saturation}. The results are obtained by fixing the parameters in Eqs. \eqref{SEt} and \eqref{CEt} and taking the late-time limit, i.e., $t\rightarrow \infty$. Note that in this limit, the only nonvanishing contribution comes from the second integrals in these equations. We note a number of key features. First, $S_E^{\rm sat.}$ increases with the dynamical exponent in both bosonic and fermionic models. Second, although in bosonic theory $C_E^{\rm sat.}$ is almost independent of the dynamical exponent, in the fermionic case the saturation value of the capacity of entanglement decreases with $z$. Moreover, while $S_E^{\rm sat.}< C_E^{\rm sat.}$ in relativistic scalar theory with $z=1$, for larger values of the dynamical exponent the entropy becomes much larger than the capacity. Thus, we expect that the corresponding reduced density matrix becomes more and more maximally mixed. This conclusion holds also in the fermionic case, where $C_E^{\rm sat.}< S_E^{\rm sat.}$ independent of $z$.

\section{Conclusions and discussions}\label{sec:diss}

In this paper, we explored the behavior of capacity of entanglement in certain nonrelativistic theories with Lifshitz scaling invariance. We considered both free-boson and free-fermion theories, which exhibit the Lifshitz scaling symmetry with a generic dynamical exponent at the UV fixed point. In order to avoid the ultraviolet divergences in the continuum limit, we regulated these models by placing them on a one-dimensional spatial lattice. Using the correlator method, we computed the capacity of entanglement in the vacuum state and investigated the dependence of its universal terms on the dynamical exponent in the specific regimes of the parameter space. We also examined the existence of a consistent entropic $c$-function based on this quantity in these nonrelativistic setups in two dimensions. Using the framework established in \cite{Alba:2017ekd}, we studied the dynamics of capacity of entanglement, which follows the quasiparticle interpretation for the spreading of entanglement. To get a better understanding of the results, we also compared these behaviors to other correlation measures, including entanglement entropy. In the following, we review our main results from these calculations, and consider their physical implications:

\begin{itemize}

\item In the ground state of a two-dimensional scalar theory with periodic boundary condition, the capacity of entanglement increased logarithmically as a function of the width of the entangling region when the mass parameter was relatively small. Further, this measure was monotonically increasing as we increased the dynamical exponent. Although $S_E<C_E$ for $z=1$, in the case of $z>1$, the entropy became much larger than the capacity. Further, we found that the entanglement entropy went on to grow indefinitely as we increased the dynamical exponent, while the capacity of entanglement grew more slowly, which seemed to be linear. Interestingly, we found that in the large $z$ limit, $C_E\ll S_E$, which showed that the corresponding reduced density matrix became more and more maximally mixed as one increased the dynamical exponent. Indeed, from Eq. \eqref{Hamildisscalar} one can consider $z$ as a measure for the effective correlation length (number of lattice sites that are correlated together), hence this behavior was consistent. Moreover, considering Dirichlet boundary condition, we found qualitatively similar results. In particular, the difference between these measures scaled logarithmically with the width of the subregion, i.e., $S_E-C_E\propto (z-1)\log \frac{\ell}{\epsilon}$, such that in the relativistic limit this universal term vanished. This observation is consistent with the previous results for two-dimensional conformal field theories, where the capacity of entanglement tracked the entanglement entropy such that at leading order, we had $C_E=S_E$.

\item In the vacuum state of a two-dimensional fermionic theory, the capacity of entanglement became completely independent of $z$ in the small mass regime. Indeed, in this case, our numerical results showed that the leading behavior of $C_E$ was logarithmic, whose coefficient coincided with the relativistic case with $z=1$. Of course, similar results have been previously found for $S_E$ in \cite{Vasli:2024mrf}. Indeed, this behavior was due to the specific structure of the higher-order spatial derivatives in action \eqref{actiondirac}. Moreover, the numerical results showed that in the small mass regime, the capacity became larger than the entropy, hence the width of the eigenvalue distribution of the corresponding reduced density matrix decreased. Further, considering a finite mass, although $S_E>C_E$ for small subregions, in the case of $\ell \gg z\epsilon$, the entropy became smaller than the capacity.

\item Examining the existence of a convenient entanglement $c$-function based on capacity of entanglement, we found that although the corresponding function decreased monotonically for the case of $z=1$, for larger values of the dynamical exponent it failed to satisfy this property. Thus, we may conclude that for $z>1$, where the Lorentz invariance was expected to break down, $c_C$ was not a genuine $c$-function. Interestingly, we found that another $c$-function based on the second moment of shifted modular Hamiltonian existed, which behaved monotonically in the same regime of the parameter space.
Let us recall that the proof of the entropic version of the $c$-theorem is based on some basic properties: the Lorentz symmetry and unitarity of the underlying field theory, which combined with the strong subadditivity of the entanglement entropy \cite{Casini:2004bw}. Hence, at first sight it is not surprising that in breaking the relativistic invariance, the corresponding $c$-function based on capacity of entanglement failed to produce the expected monotonicity. However, our results showed that even in this nonrelativistic setup with $z>1$, other entanglement-based $c$-functions existed, such as $c_S$ and $c_M$, which displayed a monotonic behavior under the RG flow.

\item Considering the dynamics of entanglement followed by a global quantum quench, we have provided a numerical analysis and examined the various regimes in the growth of capacity of entanglement using the quasiparticle picture in both bosonic and fermionic theories. In the bosonic case, an interesting observation is that when the postquench Hamiltonian was scale invariant, although the entropy density diverged in specific limits, the capacity density saturated to its maximum value, whose rate was controlled by the dynamical exponent. Thus, the production rates of quasiparticles and their individual contributions to these measures were completely different. Moreover, for generic integer values of $z$, the region in momentum space with maximum velocity did not necessarily coincide with the region where entropy and capacity densities reached their maximum. Indeed, we found that only in the relativistic case did these two regions coincide. Hence, the contribution of fast quasiparticles to $S_E(t)$ and $C_E(t)$ was dominant, and the entanglement was mainly carried by these modes. On the other hand, for larger values of the dynamical exponent, when the densities reached their maximum values, the group velocity vanished, thus most of the entanglement was carried by the slow modes. Further, the capacity density was a monotonically increasing function of the dynamical exponent and the prequench mass, as expected. Extending this analysis to the fermionic theories, the corresponding results were slightly different. A key observation to note is that although for $z=1$ the contribution of fast modes to $S_E(t)$ was dominant, their contribution to $C_E(t)$ was negligible, hence this quantity was mainly carried by the slow modes. On the other hand, for larger values of the dynamical exponent, the contribution of fast modes to the capacity of entanglement became more pronounced.

\end{itemize}

We can extend this study to different interesting directions. A simple generalization of our work would be to understand
the scaling of capacity of entanglement in higher-dimensional nonrelativistic free models, for which similar techniques used here trivially apply. A more difficult generalization would be to extend our work to more realistic models with nontrivial interactions, where we believe the entanglement measures will exhibit more interesting features. Another interesting direction is to consider symmetry-resolved capacity of entanglement, similar to \cite{Arias:2023kni,Murciano:2019wdl}. We leave the details of some interesting problems for future study.

\subsection*{Acknowledgements}
We would like to thank Krysztof Andrzejewski for valuable discussions and correspondence. 
Some numerical computations related to this work were carried out at the Department of Physics Computer
Facility. The work of M. Reza Mohammadi Mozaffar is supported by the Iran National Science Foundation (INSF) under Project No. 4036941.

\subsection*{Data availability}
The data that support the findings of this article are not publicly upon publication because it is not technically feasible and/or the cost of preparing, depositing, and hosting the data would be prohibitive within the terms of this research project. The data are available from the authors upon reasonable request.

\end{document}